# Evidence of a Two-Component Jet in the Afterglow of GRB 070419A

ZHENG Weikang[1,2†], DENG Jinsong[1]

1 National Astronomical Observatories, Chinese Academy of Sciences, Beijing 100012, China;
2 Graduate University of Chinese Academy of Science, Beijing 100049, China

A two-component jet model is proposed to explain the unusual afterglow of GRB 070419A. Regarding the optical light curve, a wide "jet" with an opening angle of > 30-40 degrees is assumed to produce the late shallow decay, while the three early power-law segments must be caused by a narrow jet with an opening angle of ~ 2-4 degrees. Additional energy injections to both components are required. Late X-ray emission may come from either the wide jet or the narrow one. If the latter is correct, the jets may run into an ISM environment and the temporal index of the late energy injection may be *q* ~ 0.65.

gamma rays: bursts, gamma rays: observations

Received; accepted
†Corresponding author (email: zwk@bao.ac.cn)
Supported by National Basic Research Program of China-973 Program 2009CB824800

Multi-wavelength observations of the afterglow of GRB 070419A ($T_{90}$ ~ 110s) have been published[1]. Its optical light curve showed complex behaviours which can be described by four power-law segments[1]. It began with a rising phase, denoted as *f*1, at *t* < 460s with a power-law index of $\alpha = -1.56 \pm 0.70$, followed by a slow decay (*f*2 phase) between 460s and 1500s with $\alpha = 0.61 \pm 0.09$, then a fast decay (*f*3 phase) with $\alpha = 1.51 \pm 0.12$, and finally another slow decay (*f*4 phase) with $\alpha = 0.41 \pm 0.17$ at $t > 10^4$s. The X-ray light curve is simpler, comprising of an early fast exponential decay ($t < 10^3$s), which must be the tail of prompt GRB emission, and a power law ($t > 10^3$s) which is the real afterglow component. X-ray data after ~ $10^3$s are sparse so two fitting values of the late-time power-law index were provided[1], which are $\alpha_X = 1.27^{+0.18}_{-0.12}$ and $\alpha_X = 0.65^{+0.35}_{-0.39}$. The spectral index of the late X-ray data is $\beta_X = 1$[1].

The complex optical light curve of the GRB 070419A afterglow is difficult to understand with the simple standard fireball forward/reverse shock model[1]. In the simple model, the relativistic outflow is assumed to be in the form of a homogeneous jet. In this paper, we examined if a model invoking two jet components can explain the abnormal afterglow light curve of the GRB.

## 1. The optical afterglow light curve

In the two-component jet model[2], the GRB outflow consists of a narrow ultra-relativisitic core jet and less relativistic surrounding material which can be regarded as a wide "jet". Of the optical afterglow of GRB 070419A, a natural explanation is that the last *f*4 phase was due to the forward shock of the wide jet while the early *f*1-*f*3 phases were dominated by emissons of the narrow jet. Following the one-component jet model[1], the light curve peak at *t* ~ 460s must be the narrow-jet deceleration time from which an initial Lorentz factor of $\Gamma_n$ ~ 350 can be derived.

Additional energy injections into the afterglow are required to explain the slow decays in the *f*2 and *f*4 phases, as in the popular refreshed shock model[3]. But we think the second light curve break at ~1500s is the jet break, rather than caused by the cessation of energy injections into the narrow jet. The latter would result in an unnatural hollow outflow structure at late time since energy injections into the wide jet are assumed to continue into the *f*4 phase. The change of the temporal index, i.e., $\Delta\alpha = 1.51 - 0.61 = 0.9$, is in agreement with a jet break. A jet break time of ~ 1500s corresponds to a jet opening angle of ~ 1.9 degrees for the ISM case and of ~ 3.8 degrees for the wind case, assuming the typical parameter values of $n_0 = 3$, $\eta_\gamma = 0.2$ and $A_* = 1$[4].

A lower limit for the opening angle of the wide jet can be estimated since no light curve break was observed in the $f4$ phase up to ~ $10^7$s. The opening angle must be larger than ~ 42 degrees for the ISM case and larger than ~ 30 degrees for the wind case.

## 2. The late-time X-ray afterglow

Unlike the $f4$ phase of the optical afterglow, the late-time X-ray emission (after ~ 2000s) may come from either the narrow or the wide jet. These two cases are discussed respectively in the following.

In the narrow jet case, during the optical $f3$ phase, both the optical photons and X-rays were of the same origin, i.e. the narrow jet after its hypothesized jet break. So the temporal index of the late X-ray light curve is more likely ~ 1.27, close to the optical value of ~ 1.5, than ~ 0.65. The temporal indices, however, are still somewhat smaller than the typical post-break decay slope of > 1.5, which could be explained if there were still energy injections into the narrow jet after the jet break. No discrimination could easily be made in this case between an ISM enviroment and a wind one.

In the wide jet case, both the optical $f4$ phase and the late X-ray emission are attributed to the wide jet with additional energy injections. We can compare the theoretical temporal and spectral indices of the refreshed shock model[3] with the observed values, which are $\alpha_X$ ~ 0.65 or 1.27, $\alpha_{opt} = 0.41 \pm 0.17$, and $\beta_X = 1$.

If $\alpha_X$ ~ 0.65, we can identify two possible theoretical cases[3]. One is slow-cooling in an ISM environment with $\nu_m < \nu_{opt} < \nu_c$ and $\nu_X > \nu_c$. First we have $p = 2$ since $\beta_X = 1 = p/2$. The closure relation for the X-ray band, i.e., $\alpha = (q-2)/2 + (2+q)\beta/2$, gives $q = 0.65$ for the temporal index of the energy injection. Then the model slope of the optical light curve is $\alpha = [(2p-6) + (p+3)q]/2 = 0.31$, which is consistent with the observed $\alpha_{opt} = 0.41 \pm 0.17$. The other case, i.e., fast-cooling in a wind environment with $\nu_{opt} < \nu_c$ and $\nu_X > \nu_m$, can actually be excluded. This is because an afterglow always evolves from fast cooling to slow cooling: $\nu_m$ decreases with time, while in a wind environment the cooling frequency $\nu_c$ increases with time. Unreasonable physical parameters may be needed to keep $\nu_c < \nu_m$ as late as up to ~ $10^6$s.

For the wide jet case, $\alpha_X$ ~ 1.27 can not be the correct value. Here the only possibility would be slow cooling in a wind environment with $\nu_{opt} < \nu_m$ and $\nu_m < \nu_X < \nu_c$. The closure relation in the X-ray band would give $q = 0.27$. The corresponding optical temporal index then would be $\alpha = (q-1)/3 = -0.24$, which means a rising light curve instead of the observed decay one.

## 3. Discussions

Recently a two-component jet model has also been applied to GRB 080319B which had a bright, naked-eye optical afterglow[5]. In both GRBs, the telltale feature of two jet components is a late slow decay phase in the afterglow light curve following an earlier shallow-to-steep break. In GRB 070419A such a feature was seen in the optical band, while in GRB 080319B the X-ray band. The two-componet jet model of GRB 080319B is more reliable since a very-late X-ray light curve break was detected, which can be modelled as the jet break of the wide component.

In the one-component jet model, the scenario that a reverse shock shaped the early optical light curve ($f1$-$f3$ phases) can be rejected in the ground that the predicted increasing emission of the forward shock until a few $10^6$s was not observed[3]. This argument is no longer valid for the two-component jet model since the observed $f4$ phase came from a wide jet and hence independent of the $f1$-$f3$ phases of a narrow jet origin. One can assume that the forward shock emission of the narrow jet was overwhelmed by the that of the wide jet. The observed temporal index of ~ 1.51 in the $f3$ phase is closer to the value predicted by the reverse shock theory for an ISM environment (~ 1.9) than that for a wind environment (~ 3). The $f2$ phase with $\alpha$ ~ 0.61 can be explained if the characteristic frequency $\nu_{m,r}$ was above the optical band and the break at ~ 1500s was due to $\nu_{m,r}$ crossing the optical band, corresponding to the theoretical case of $\mathscr{R}_\nu < 1$ [7].

*The authors are grateful to X.F. Wu and S. Kobayashi for stimulating discussions and comments. This work has been supported by National Basic Research Program of China – 973 Program 2009CB824800 and by National Natural Science Foundation of China – Grant No. 10673014 and No. 10873017.*